\documentclass[twocolumn,superscriptaddress,pra,amsfonts,showpacs]{revtex4}

\RequirePackage[utf8]{inputenc}
\RequirePackage[T1]{fontenc}

\RequirePackage[intlimits]{amsmath}
\RequirePackage{amssymb,amstext,color,xspace}
\RequirePackage{soul}

\providecommand\x{\mathbf{r}}
\providecommand\kv{\mathbf{k}}
\providecommand\DV{\mathbf{D}}
\providecommand\EV{\mathbf{E}}
\providecommand\PV{\mathbf{P}}
\providecommand\AV{\mathbf{A}}

\providecommand\Fcal{\mathcal{F}}

\providecommand{\ket}[1]{\left\vert #1\right\rangle}
\providecommand{\bra}[1]{\left\langle #1\right\vert}

\newcommand\conditionToSatisfy{\ensuremath{\overset{!}{=}}}

\newcommand{\kM}{k_\text{M}}

\begin{document}

\title{Fundamental limitation of ultrastrong coupling between light and atoms
}

\newcommand{\ourAddress}{\affiliation{Institute for Solid State Physics and Optics, Wigner Research Centre for Physics,\\Hungarian Academy of Sciences, P.O. Box 49, H-1525 Budapest, Hungary}}

\author{András Vukics}
\email{vukics.andras@wigner.mta.hu}
\ourAddress
\author{Tobias Grießer}
\ourAddress
\author{Peter Domokos}
\ourAddress

\begin{abstract}\noindent In a recent work of ours \cite{vukics14}, we generalized the Power–Zineau–Woolley gauge to describe the electrodynamics of atoms in an arbitrary confined geometry. Here we complement the theory by proposing a tractable form of the polarization field to represent atomic material with well-defined intra-atomic potential. The direct electrostatic dipole-dipole interaction between the atoms is cancelled. This theory yields a suitable framework to determine limitations on the light-matter coupling in quantum optical models with discernible atoms. We find that the superradiant criticality is at the border of covalent molecule formation and crystallization.
\end{abstract}


\maketitle

\section{Introduction}
\label{sec:intro}
Ultrastrong coupling between two systems is realized when the characteristic coupling constant of the interaction is of the same order of magnitude as the bare frequencies of the systems without interaction. Although this definition is meaningful regardless of the type of interaction, in reality it is only the electromagnetic interaction that allows for such level of control on the quantum scale as is necessary for any chance at ultrastrong coupling. Reaching the ultrastrong-coupling regime is an outstanding objective in controlled laboratory systems where confined electromagnetic radiation interacts with some kind of material degree of freedom \cite{Guenter09,baumann2010dicke,niemczyk2010circuit,forn2010observation,todorov2010ultrastrong,Scalari2012Ultrastrong,geiser2012ultrastrong,Todorov12,Todorov14a,Todorov14b}. It would mean the realization of hybrid light-matter “molecules”. On one hand, the essential hybridization of electronic states by photons leads to fundamental effects, such as the dynamical Casimir effect \cite{Ciuti2005,johansson2009dynamical,wilson2011observation} or quantum criticality \cite{Hepp73,Wang73,Emary03a,Zou14}. On the other hand, the external control over the properties of such artificial objects opens the way to  novel electro-optical applications where the “light-matter molecule” serves as a quantum interface between degrees of freedom of completely different kinds. The possibility of reaching an interaction of such strength touches unavoidably upon the foundations of the electromagnetic interaction in those system (quantum electrodynamics – QED). In the present paper, we aim at theoretically clarifying the grounds for ultrastrong coupling of light with atoms. 

In our recent work \cite{vukics14}, we generalized the Power–Zineau–Woolley (PZW)  gauge \cite{power1959coulomb,woolley1971molecular,power1978nature} (cf. also Section IV.C in \cite{CDG}) to arbitrary confined geometry, which includes any cavity QED situation, and is hence of fundamental importance for defining single- or few-mode models that are widely used in quantum optics. We argued that this picture is better suited as a basis of such models than the Coulomb gauge, because of the lack of direct coupling between the modes (which is present in the Coulomb gauge through the A-square term), and the lack of direct, electrostatic atom-atom interactions (which is also present in the Coulomb gauge). Indeed, a term-by-term correspondence between the full Hamiltonian in the new picture and standard model Hamiltonians of quantum optics (Jaynes–Cummings or Rabi models) was found \cite{Keeling07,Vukics12}. An immediate consequence of using this gauge is that the feasibility of the superradiant phase transition in the Dicke model cannot be excluded  by the commonly known no-go theorem based on the A-square term \cite{Rzazewski75,Nataf10}. However, the evaluation of the accessible coupling strength requires the clarification of the polarization field used to describe the material component in the PZW gauge. This is the subject of the present paper for the case of a medium composed of isolated atoms.

The single-atom Hamiltonian in the PZW gauge (also called multipolar gauge) can be expressed as
\begin{equation}
\label{eq:SingleAtomNew}
H_A=\sum_{\alpha\in{A}}\frac{\mathbf{p}_\alpha^2}{2m_\alpha}+\frac1{2\varepsilon_0}\int_{\text{supp}(\PV_A)}d^3r\,\PV_A^2,
\end{equation}
where $A$ denotes a single atom, $\alpha$ labels the constituent point charges, $\mathbf{p}_\alpha$ is the momentum of the point charge in the new picture, and $\PV_A$ is the polarization field associated with the atom. The second term in this Hamiltonian, which can be viewed as the ``potential'' term, can be  problematic. This is the case for the intuitive choice of $\PV_A$, known as the Power form (Section IV.C.1.a in \cite{CDG}), which is directly based on the point-charge distribution within the atom:
\begin{equation}
\label{eq:PowerP}
 \PV_{\text{Power},A}(\x)=\sum_{\alpha\in A} q_\alpha\x_{\alpha}\int_0^1 du\,\delta(\x-u\x_\alpha),
\end{equation}
where $q_\alpha$ is the charge and $\x_\alpha$ is its position. This form is a distribution, so that (\ref{eq:SingleAtomNew}) contains a distribution squared, that is mathematically ill defined. In this paper we present an alternative definition of the polarization field which gives a meaningful ``potential term'' in the single-atom Hamiltonian and, at the same time, retains the good property that there is no direct electrostatic  interaction between different atoms. The proposed definition allows thus for exploring the limits of ultrastrong coupling between atomic matter and light. 

The discussion is structured in the following way. We start, in Section \ref{sec:overview}, with pointing out that the definition of the PZW canonical transformation has a freedom, namely, the freedom of choosing the transverse part of the polarization field. We identify the restrictions this field has to meet. In Section \ref{sec:choiceOfP_Trans}, using this freedom, we define the transverse part of the polarization field generated by spatially distinct, well-localized, neutral charge clusters (atoms). With the proposed definition, the passage to the multipolar gauge is in fact a class of transformations parametrized by a single wavenumber-cutoff parameter. We show how to define bounds for the cutoff wavenumber on intuitive physical basis. The lower comes from the requirement of the disappearance of the A-square term (Section \ref{sec:lowerLimit}), while the upper from that the potential term in Hamiltonian (\ref{eq:SingleAtomNew}) be only a slightly perturbed Coulomb potential (Section \ref{sec:upperLimit}).

With this appropriate choice, in Section \ref{sec:limits}, we arrive at a picture where the atoms interact exclusively via the radiation modes of the electromagnetic field by means of photon emission and absorption processes. Since the instantaneous electrostatic dipole-dipole Coulomb interaction between spatially distinct atoms is canceled, the atoms have well-defined resonances regardless of the presence of other atoms. Thus, the commonly used few-mode models, such as the Dicke model, can be straightforwardly defined. The necessary requirement is that the atoms do not approach each other within the distance corresponding to the cutoff, i.e., they all are surrounded by an ``intimacy zone''.  Therefore, on the other hand, the proposed gauge is unsuited for describing molecule formation or solidification. This is to be noted because our results indicate that the superradiant quantum phase transition accompanied by ferroelectric ordering of the atoms in the dipolar Hamiltonian (most simply in the Dicke model) blends into that commonly observed phase transition which is solidification. Nevertheless, sufficiently far from the regime of molecule formation, the proposed gauge can be used to evaluate the limitation of ultrastrong coupling within the Dicke and related models.

\section{Quantum electrodynamics of atoms in the multipolar gauge}
\label{sec:overview}

The Hamiltonian in the multipolar gauge reads \cite{vukics14}
\begin{equation}
\label{eq:Hamiltonian}
H= H_\text{EM} + H_\text{kin} + \frac1{2\varepsilon_0}\int dV\,P^2 - \frac1{\varepsilon_0}\int dV\,\PV\cdot\DV,
\end{equation}
where $H_\text{EM}$ is the Hamiltonian of the free electric field, $H_\text{kin}$ is the kinetic energy of the atoms, and $\DV$ is the (purely transverse) electric displacement field. $\PV$ is the polarization field, which plays a crucial role in this gauge. This satisfies the relation
\begin{equation}
\label{eq:PlongFromCharges}
 \nabla\cdot\PV=-\rho,
\end{equation}
that connects the polarization field to the charges, however, since $\nabla\cdot\PV=\nabla\cdot\PV^\Vert$, it relates \emph{only} to the longitudinal component of the field \footnote{Here, by longitudinal we mean that the field can be written as gradient of such a scalar field as vanishes on the eventual boundaries. Conversely, a transverse field is divergence-free and normal to the boundaries. For a general definition and elaborate treatment of these types of fields cf. \cite{vukics14}. Using the formalism of that paper, our present treatment that for the sake of transparency is restricted to free space, can be straightforwardly extended to arbitrary confined geometry.}. Equivalently:
\begin{equation}
\label{eq:PlongFromElectricField}
 \PV^\Vert=-\varepsilon_0\EV^\Vert,
\end{equation}
meaning that the part of the electric field which is attached to the charges and follows their motion instantaneously (the so-called ``near field''), is incorporated into the polarization field, that represents the material component of the interacting system.

Besides $\PV^\Vert$, the other orthogonal component of $\PV$ is the transverse component, $\PV^\bot$. Transverse fields are source-free and normal to boundaries, that is, they can either be written as curl of vector fields normal to boundaries, or are cohomological \cite{vukics14}. We emphasize again that the transverse component of the polarization field \emph{is not determined by the charges} in such a direct way as Eq.~(\ref{eq:PlongFromCharges}) for the longitudinal one. Instead, it has to obey a set of conditions in order that the multipolar gauge be really useful. The first comes from the requirement of the “elimination of the A-square term” to electric-dipole order (see Eq. (5c) in Ref.~\cite{vukics14})  that we expect from this gauge:
\newcounter{equationbefore}\setcounter{equationbefore}{\value{equation}}\setcounter{equation}{0}
\renewcommand{\theequation}{Cond. \Roman{equation}}
\begin{equation}
\label{cond:Asquare}
\frac\partial{\partial\x_\alpha}\int dV\,\PV^\bot\cdot\AV\conditionToSatisfy q_\alpha\,\AV(\x_A),\quad\forall\alpha\in A,
\end{equation}
where $\x_A$ is the position of that atom $A$ to which the particle $\alpha$ with position $\x_\alpha$ and charge $q_\alpha$, belongs. This makes the kinetic momentum coincide with the canonical one up to a magnetic term (Röntgen term, cf.~Section IV.C.4.c in \cite{CDG}), which we neglect here \footnote{Actually, (\ref{cond:Asquare}) eliminates the A-square term only in a long-wavelength approximation because the condition for exact cancellation would read (cf. \cite{vukics14} Eq. (15))\[\frac\partial{\partial\x_\alpha}\int dV\,\PV^\bot\cdot\AV=q_\alpha\,\AV(\x_\alpha).\] This condition cannot be met \emph{exactly} because the LHS, being a gradient of a scalar field, is curlfree (meaning curl with respect to $\x_\alpha$), but the RHS is not. This argument is removed by the long-wavelength approximation because in (\ref{cond:Asquare}) both sides have vanishing curl with respect to $\x_\alpha$.}.

The next condition is necessary for eliminating the electrostatic dipole-dipole interaction between the atoms in this gauge, which means that the medium can be considered as independent atoms (elimination of cross-coupling in the P-square term). It requires that $\PV_A$, the polarization field corresponding to any single atom $A$ (so that the full polarization is $\PV=\sum_A\PV_A$), have finite support around the atomic position. The size of the support can be a parameter that for instance will play an important role with our choice of $\PV^\bot$ in Section~\ref{sec:choiceOfP_Trans}. This fact of finite support we express here with the help of the “long-wavelength” delta function as:
\begin{equation}
\label{cond:atomicDipoles}
 \PV_A(\x)\simeq\mathbf{d}_A\delta_<(\x-\x_A).
\end{equation}
Through the example of our choice of $\PV^\bot$, in Section~\ref{sec:choiceOfP_Trans} we will discuss why and in what sense the equality is only approximate here. For the moment it is sufficient that the field $\PV_A$ should describe a well-localized dipole when regarded on a lengthscale much larger than the atomic size.

Under this last condition, and assuming the atoms distant enough from each other so that the supports of their respective $\PV_A$ fields do not overlap, the third term of the Hamiltonian~(\ref{eq:Hamiltonian}) – the potential term – can be written as per atom:
\newcounter{conditionbefore}\setcounter{conditionbefore}{\value{equation}}
\renewcommand{\theequation}{\arabic{equation}}
\setcounter{equation}{\value{equationbefore}}
\begin{equation}
U\equiv\frac1{2\varepsilon_0}\int dV\,P^2 =\frac1{2\varepsilon_0}\sum_A \int dV\,P_A^2\equiv\sum_AU_A,
\end{equation}
which means that the only interaction between different atoms is the indirect one via the displacement field $\DV$, that involves the emission and absorption of photons, included in the Hamiltonian~(\ref{eq:Hamiltonian}) by its last term. This leads us to the final condition that $\PV^\bot_A$ has to satisfy for each atom $A$. The potential corresponding to atom $A$ can be separated into parts generated by the longitudinal and the transverse part of the polarization field, where the longitudinal part can be identified (cf. Eq.~(\ref{eq:PlongFromElectricField}), and Section I.B.5.a in \cite{CDG}) as the Coulomb potential term:
\begin{multline}
\label{eq:perturbationDefined}
 U_A=\frac1{2\varepsilon_0}\int dV\left(P_A^\Vert\right)^2+\frac1{2\varepsilon_0}\int dV\left(P_A^\bot\right)^2\\\equiv U_A^\text{Coul}+\Delta U_A.
\end{multline}
Then, the last condition is that the potential generated by the transverse part be only a perturbation to the normal Coulomb potential, because we do not want to upset atomic physics as it has been worked out over the last century based in leading order on the electrons experiencing the Coulomb potential as they orbit the nucleus.
\setcounter{equationbefore}{\value{equation}}\setcounter{equation}{\value{conditionbefore}}
\renewcommand{\theequation}{Cond. \Roman{equation}}
\begin{equation}
\label{cond:perturbation}
\Delta U_A\text{ is only a perturbation to }U_A^\text{Coul}.
\end{equation}

In Section \ref{sec:limits}, we show how to proceed under this set of conditions to the derivation of the standard models of quantum optics.

\section{Appropriate choice of the transverse polarization in dipole order}
\label{sec:choiceOfP_Trans}

The following form for the transverse part of the polarization field we demonstrate in the following to optimally fulfill the set conditions:
\setcounter{conditionbefore}{\value{equation}}
\renewcommand{\theequation}{\arabic{equation}}
\setcounter{equation}{\value{equationbefore}}
\begin{subequations}
\label{eq:lwDeltaFunction}
\begin{equation}
\label{eq:choiceOfS_freeSpace}
 \PV_A^\bot(\x)=\boldsymbol{\delta}_<^\bot(\x-\x_A)\,\mathbf{d}_A,
\end{equation}
where $\mathbf{d}_A=\sum_{\alpha\in A}q_\alpha\x_\alpha$ is the dipole moment of atom $A$, and we have used the long-wavelength part of the transverse delta function, most conveniently defined by a Lorentzian cutoff in k-space at the cutoff wavenumber $\kM$:
\begin{equation}
 \widetilde{\boldsymbol{\delta}}_<^\bot(\kv)=\frac1{(2\pi)^{\frac32}}\left(\mathbf{id}_{\mathbb{R}^3}-\frac{\kv\circ\kv}{k^2}\right)\frac{\kM^2}{k^2+\kM^2},
\end{equation}
where we have assumed \(\x_A=0\) for the single atom \(A\) that we are going to consider henceforth in this Section. Using the real-space form of this, we get for $r\gtrsim\kM^{-1}$
\begin{multline}
\PV_A^\bot(\x)=\frac{\eta(r)}{4\pi r^3}\left[\frac{3\left(\x\cdot\mathbf{d}_A\right)\x}{r^2}-\mathbf{d}_A\right],\\\text{with}\quad\eta(r)=1-\left(1+\kM r+\frac{\kM^2r^2}2\right) e^{-\kM r}.
\end{multline}
\end{subequations}

As it is apparent, this is just the electric field of a dipole, so that this choice of $\PV_A^\bot$  cancels the dipole order of the longitudinal component (\ref{eq:PlongFromElectricField}), outside a distance $\sim\kM^{-1}$. This makes that the support of $\PV_A=\PV_A^\Vert+\PV_A^\bot$ is indeed finite in dipole order, that is, the atoms interact only in quadrupole order, which we are neglecting here. Hence, (\ref{cond:atomicDipoles}) is fulfilled up to dipole order, which we consider sufficient for our purposes here, and is in accordance with the neglect of the Röntgen term above \footnote{This is because for a hydrogen-like atom, both this term and the quadrupole term of the electrostatic interaction is $\mathcal{O}([a_0k_\text{radiation}]^2)$, while the dipole term would be $\mathcal{O}(a_0k_\text{radiation})$, $a_0$ being the Bohr radius (cf.~Section \ref{sec:upperLimit}), for which $a_0k_\text{radiation}\ll1$ holds.}. This is the sense in which the equality in (\ref{cond:atomicDipoles}) is fulfilled only approximately: up to dipole order and without a region of $\kM^{-1}$ around the atom. This latter is the \emph{intimacy region} of an atom: the interaction between different atoms simplifies substantially to the indirect interaction mediated by the radiation field modes only if they do not penetrate each other's intimacy region.

This is also a point where our approach manifestly deviates from the original Power–Zineau–Woolley approach, since with Power’s original definition of the polarization field (\ref{eq:PowerP}), it exactly vanishes outside of zero-measure regions within the atom (on the other hand, (\ref{cond:perturbation}) is impossible to fulfill with that choice of the polarization field).

In our case, the cutoff wavenumber $\kM$ is not a renormalization parameter, but parametrizes a class of allowed transformations. In the following, we use the remaining two conditions (\ref{cond:Asquare}) and (\ref{cond:perturbation}) to define an interval for $\kM$. 

\subsection{Lower limit of cutoff}
\label{sec:lowerLimit}

The lower limit comes from (\ref{cond:Asquare}) as follows. Taking the free-space traveling-wave expansion of the vector potential $\AV(\x)=\int d^3k\sum_{\epsilon}
 \left[\boldsymbol\alpha_{\epsilon}(\kv)\,e^{i\kv\cdot\x}+\text{c.c.}\right]$, and substituting into the LHS of (\ref{cond:Asquare}) with our choice of $\PV^\bot$ (\ref{eq:choiceOfS_freeSpace}), we obtain
\begin{multline}
 \int dV\,\frac{\partial\PV^\bot}{\partial\x_\alpha}\AV=q_\alpha\int d^3k\,\frac{\kM^2}{k^2+\kM^2}\sum_{\epsilon}
 \left[\boldsymbol\alpha_{\epsilon}(\kv)+\text{c.c.}\right]\\\equiv q_\alpha\AV_<(0).
\end{multline}
That is, if $\boldsymbol\alpha_{\epsilon}(\kv)\approx0$ for $k\geq\kM$ in the course of the dynamics, so that we can write $\AV_<(0)=\AV(0)$, then (\ref{cond:Asquare}) is fulfilled. Hence the lower limit for $\kM$:
\setcounter{equationbefore}{\value{equation}}\setcounter{equation}{0}
\renewcommand{\theequation}{$\kM$ lower limit}
\begin{equation}
\label{eq:lowerLimit}
 \kM\gg k_\text{radiation},
\end{equation}
In simple terms, the atom has to be small compared to the wavelength of populated modes, represented by $k^{-1}_\text{radiation}$. This is usually termed the dipole approximation, or, more precisely, the long-wavelength approximation. In the case of a hydrogen atom and optical frequencies, there are roughly 3 orders of magnitude between the atomic size and the characteristic wavenumber of the atomic transitions. This can be seen by expressing the characteristic transition frequency with fundamental constants as $\hbar\omega_\text{A}=\frac38 m_\text{e}c^2\alpha^2$, and using $m_\text{e}c\alpha=\hbar/a_0$ to obtain
\renewcommand{\theequation}{\arabic{equation}}
\setcounter{equation}{\value{equationbefore}}
\begin{equation}
 k_\text{A}=\frac{3\alpha}8 \frac 1{a_0},
\end{equation}
where $\alpha$ is the fine-structure constant, and $k_\text{A}$ determines the wavenumber of the relevant resonant radiation modes ($k_\text{radiation}\sim k_\text{A}$).

The long-wavelength approximation is an elementary requirement for atom optics, because in this field we need well-defined atoms with level structures determined by a static potential, and radiative effects on the atomic scale are treated as mere perturbations.

\subsection{Upper limit of cutoff}
\label{sec:upperLimit}

The upper limit comes from (\ref{cond:perturbation}) as follows. With our choice of the transverse polarization field (\ref{eq:choiceOfS_freeSpace}), we can take the potential resulting from $\PV^\bot$ seriously as a physically significant potential. This is in sharp contrast to Power’s choice, since in that case the “perturbation” is infinite and calls for renormalization. The finite and regular potential $\Delta U_A$ in Eq.~(\ref{eq:perturbationDefined}) resulting from the proposed polarization field $\PV^\bot$ is
\begin{equation}
 \Delta U_A=\frac1{2\varepsilon_0}\int d^3k\,\left|\widetilde{\PV}^\bot_A\right|^2=\frac{\kM^3}{24\pi\varepsilon_0}d_A^2.
\end{equation}
This potential term manifests that the atom is defined in the multipolar gauge differently from the one in Coulomb gauge, i.e.,  the gauge transformation shifts the boundary between atom and field.

To keep our discussion as simple as possible, the effect of this potential will be calculated in the example of the 1s state of the hydrogen atom by time-independent perturbation theory. To be able to do quantum physics, the potential has to be treated as an operator acting on the Hilbert space of the atom’s constituents. Here, this consists in considering the positions of the atomic constituents appearing in \(\mathbf{d}_A\) as quantum operators. For the hydrogen atom, \(\mathbf{d}=e\,\x\), where $\x$ is the relative coordinate, so that the perturbing potential reads
\begin{equation}
 \Delta U_\text{hydrogen}=\frac{e^2\kM^3}{24\pi\varepsilon_0}\,r^2.
\end{equation}
The 1s wave function of the electron reads $ \Psi_\text{1s}(\x)=e^{-r/a_0}/\left(\sqrt\pi a_0^{3/2}\right)$, where \(a_0\) is the Bohr radius. The first-order perturbation reads:
\begin{multline}
 E_\text{1s}^{(1)}=\bra{\text{1s}}\Delta U_\text{hydrogen}\ket{\text{1s}}\\=\frac{e^2\kM^3}{24\pi^2\varepsilon_0 a_0^3}\int d^3r\,r^2 e^{-\frac{2r}{a_0}}=\frac{e^2\kM^3a_0^2}{8\pi\varepsilon_0}.
\end{multline}
Let us compare this energy to the binding energy of the hydrogen atom, the Rydberg energy. We find the following remarkably simple expression:
\begin{equation}
\label{eq:energyRatio}
\frac{E_\text{1s}^{(1)}}{\text{Ry}}=\left(\kM a_0\right)^3.
\end{equation}
Hence (\ref{cond:perturbation}) is translated to an upper limit of the cutoff wavenumber as
\setcounter{equationbefore}{\value{equation}}\setcounter{equation}{0}
\renewcommand{\theequation}{$\kM$ upper limit}
\begin{equation}
\label{eq:upperLimit}
 \kM^3\ll (a_0)^{-3}.
\end{equation}
The cubic power in this expression makes that e.g. a cutoff of \(\kM\approx 1/(2\,a_0)\) already gives about \(0.1\) for the energy ratio~(\ref{eq:energyRatio}). That is, an intimacy region compressed to nearly the atomic size can be chosen without significantly altering the usual Coulomb-potential form of the  atomic Hamilton operator. 
We note  that the Lamb shift is of the same order of magnitude as $\Delta U_A$ \cite{power1959coulomb,milonni1976semiclassical}, but it should be treated separately, since it comes from the \emph{interaction with the vacuum field,} that is, the last term of Hamiltonian (\ref{eq:Hamiltonian}).

\section{The limits of coupling strength}
\label{sec:limits}

With the transverse polarization fulfilling all the conditions considered above, the Hamiltonian~(\ref{eq:Hamiltonian}) is greatly simplified in the electric dipole approximation to read
\renewcommand{\theequation}{\arabic{equation}}
\setcounter{equation}{\value{equationbefore}}
\begin{multline}
\label{eq:HamiltonianED}
H=H_\text{EM}+\sum_A\left( H_A -\mathbf{d}_A\cdot\frac{\DV(\x_A)}{\varepsilon_0}\right),\\\text{ with }H_A=\sum_{\alpha\in{A}}\frac{p_\alpha^2}{2m_\alpha}+U_A,
\end{multline}
where the replacement of the interaction term (last term of Hamiltonian~(\ref{eq:Hamiltonian})) with the last term of $H$ here is justified under~(\ref{cond:atomicDipoles}) in the long-wavelength approximation for the displacement field equivalent to the one discussed in Section \ref{sec:lowerLimit}.

The significance of this form is that the standard models of quantum optics can be derived from here through two well-established approximations: (i) the single-mode approximation for the field, when the field is described as a single harmonic oscillator mode with bosonic operator $a$; and (ii) the two-level approximation for the atoms, when the atomic ensemble can be treated as a spin $N/2$, with spin operators $S$, $S_x$, etc. This yields models like the Dicke and (with an additional rotating-wave approximation) the Tavis–Cummings, whose respective Hamiltonians correspond term by term to the Hamiltonian~(\ref{eq:HamiltonianED}):
\begin{equation}
\label{eq:standardModels}
\begin{Bmatrix}H_{\text{Dicke}}\\H_{\text{Tavis–Cummings}}\end{Bmatrix}=  \omega\,a^\dagger a+ \omega_{\text{A}} S_z + \begin{Bmatrix}\frac g{\sqrt{N}} \left( a+a^\dagger \right) S_x\\\frac g{\sqrt{N}}\left( a\,S^\dagger+a^\dagger\,S\right)\end{Bmatrix}.
\end{equation}
The important point here is that no \emph{term} of Hamiltonian~(\ref{eq:HamiltonianED}) had to be neglected, so that the A-square (direct coupling between the modes) or P-square problems (direct electrostatic coupling between the atoms) \cite{Bamba14} do not appear in this treatment. This is in sharp contrast to a treatment based on the Coulomb gauge, where e.g. the A-square problem initiated an intense debate concerning the qualitative validity of the Dicke model \cite{Rzazewski75,Nataf10,vukics14,Keeling07}.

We are now in the position of addressing the question what is the maximum coupling strength achievable between the atomic medium and a radiation field mode. The limitation arises from the requirement that the interatomic distance in the many-atom ensemble must respect the intimacy region of the individual atoms. 

 
For the collective coupling of atoms to light in the models (\ref{eq:standardModels}), the figure of merit pertaining to ultrastrong coupling is the ratio
\begin{equation}
\label{eq:figureOfMerit}
\Fcal\equiv\frac{N g^2}{\omega\,\omega_\text{A}},
\end{equation}
where $N$ is the number of atoms, $\omega$ is the frequency of a radiation mode, $\omega_\text{A}$ is the resonance frequency associated with a relevant electronic transition within an atom, and $g$ is the electric-dipole coupling constant, cf. Eq. (\ref{eq:standardModels}). For example, $\Fcal=1$ corresponds to the critical point of the Dicke model, predicting a superradiant phase transition into such a ground state of the system as features spontaneous polarization \cite{Hepp73,Wang73}. The coupling $g$ depends on the mode geometry and the oscillator strength of the atomic transition. In the QED picture used here, its square can be written the same way as in the electric-dipole gauge of QED: $g^2={\omega\, d^2}/{(2 \hbar \varepsilon_0 V)}$, where $V$ is the mode volume and $d$ is the dipole moment of the atomic transition along the mode polarization. Note that the use of all the microscopic parameters ($d$, $\omega_\text{A}$, etc.) is justified here by the lack of cross-coupling in the ``P-square term'' of Eq. (\ref{eq:Hamiltonian}): without electrostatic atom-atom interaction, the presence of other atoms does not alter the atomic level structure.

The figure of merit can be expressed in two instructive forms. Firstly:
\begin{subequations}
\label{eq:naiveArgument}
\begin{equation}
\label{eq:naiveArgumentOne}
\Fcal=\frac{N}{V} \, \lambda_\text{A}^3\,  \frac{3}{8\pi^2} \, \frac{1}{Q},
\end{equation}
where we use the quality factor of the atomic resonance $Q\equiv\omega_\text{A}/\gamma$ with $\gamma$ being the linewidth (half width at half maximum) and $\lambda_\text{A}$ the wavelength. This form expresses that the density $N/V$ must be large to compensate for the large quality factor of an atomic resonance in the denominator (for alkali atoms this is $\sim1.2-1.5\cdot10^8$). In order to get deeper insight into scaling laws, we consider transitions between hydrogen-like ground and excited states, and find the second form:
\begin{equation}
\label{eq:naiveArgumentTwo}
\Fcal = \frac{N}{V} 16\pi a_0^3\, ,
\end{equation}
\end{subequations}
where $a_0$ is the Bohr radius. That is, in order to have  $\Fcal\sim 1$,  the atomic medium should be so dense that one atom occurs per about 4 Bohr radius cube. 

This density is noticeably close to the limit allowed by the maximum of the cutoff parameter $\kM$ given by  (\ref{eq:upperLimit}), i.e., \(\kM^{-1}=2\,a_0\), indicating an interatomic distance just around 4 Bohr radius. This means that we can approach the critical-coupling point in the ultrastrong regime with an ensemble of independent atoms, i.e., \(\Fcal\lesssim1\). However, this happens at the density when the intimacy regions of adjacent atoms touch. Further increasing the density, the independent-atom model breaks down and electrostatic interactions become relevant.  Moreover, at such a small atomic distance, the overlap between the exponentially decaying electron wavefunctions ($\propto e^{-r/a_0}$) belonging to different atoms becomes significant. Such a delocalization of electronic orbits would lead to covalent bond between atoms, but this effect is not described by our model, as here the electrons do not experience the electrostatic potential of other nuclei. The role of electron-orbit delocalization at such high densities can be revealed by comparing the critical density $7\cdot10^{27}/\text{m}^3$ for Rubidium (calculated from Eq.~(\ref{eq:naiveArgumentOne}) by putting $\Fcal=1$) to the crystalline density $11\cdot10^{27}/\text{m}^3$. We find similar correspondencies also for other species used in atomic optics experiments. One can thus conjecture that the superradiant criticality blends into the commonly known criticality of solidification \footnote{In fact, the connection between superradiant criticality and solidification is somewhat more than a conjecture, since it can be supported by a scaling argument: As it is apparent in Eq.~(\ref{eq:naiveArgumentTwo}), \emph{in the dipolar picture,}  the figure of merit depends only on atomic parameters, as the mode frequency drops out from the expression of $\Fcal$ (this would not happen in Coulomb gauge). This means that the Dicke critical point ($\Fcal=1$) is determined by the Bohr radius (or, for alkaline atoms, the corresponding atomic size), $a_0$. In an atom-picture based on the Coulomb potential, which is scale-less, there is only this single atomic lenghtscale, which determines also the covalent-bond distance. As there is no free parameter for nature to separately tune the Dicke critical and the solidification point, the similarity of the two is not a mere coincidence.}.

\section{Conclusions}

We considered the quantum electrodynamic foundations of the interaction of atoms with light. In particular, we analysed in detail how the atomic medium can be defined in terms of a polarization field such that the atom-atom interactions simplify to an indirect coupling via the radiation field modes. The main difference of our approach compared to the usual treatment of the PZW transformation is that while $\PV^\Vert$ is kept exactly to represent the charge distribution via Eq.~(\ref{eq:PlongFromCharges}), we acknowledge that we have a freedom in choosing the transverse part of the polarization field. We determined the full set of  conditions (\ref{cond:Asquare} – \ref{cond:perturbation}) that the transverse polarization field in the Power–Zineau–Woolley canonical transformation has to obey.

With this background, we identified an upper limit on the coupling between atomic medium and radiation. This limit comes close to criticality in the ultrastrong coupling regime, but at this point such physical effects start to play a role as molecule formation and solidification, that rule out the treatment of matter as independent electric dipoles. This also shows that – barring artificial open systems like circuit cavity QED or such special model systems of e.g. the Dicke model as in Ref.~\cite{baumann2010dicke}, that is, systems not based on a gas of dipoles interacting with the electromagnetic field – the ultrastrong critical coupling can eventually be reached by such polarizable medium as is compacted into a solid.

As an outlook, we note that to go beyond the limit we concluded, one should abandon the isolated-atom picture, and allow the atoms to penetrate each other’s intimacy zone. This situation could be studied by a many-body model with a hard-core inter-atomic potential. The determination of the form of this potential and the study of such a many-body model is currently under way by the authors.

\section*{Acknowledgments}

We thank the anonymous Referee of our last submission for the idea of the scaling argument presented in footnote [32]. This work was supported by the Hungarian Academy of Sciences (Lend\"ulet Program, LP2011-016). A.V. acknowledges support from the János Bolyai Research Scholarship of the Hungarian Academy of Sciences.


\end{document}